\documentclass[aps,reprint]{revtex4-1}
\usepackage{amsmath,amssymb,graphicx}

\newcommand{\bra}[1]{\ensuremath{\left\langle {#1} \right|}}
\newcommand{\ket}[1]{\ensuremath{\left|  #1 \right\rangle}}

\draft 

\begin{document}

\title{Reducing Collective Quantum State Rotation Errors with Reversible Dephasing} 



\author{Kevin C. Cox}
\author{Matthew A. Norcia}
\author{Joshua M. Weiner}
\affiliation{JILA, NIST, and University of Colorado, 440 UCB, 
Boulder, CO  80309, USA}
\author{Justin G. Bohnet}
\altaffiliation{present address: National Institute of Standards and Technology, Boulder, Colorado 80305, USA}
\author{James K. Thompson}
\affiliation{JILA, NIST, and University of Colorado, 440 UCB, 
Boulder, CO  80309, USA}
\email[]{kevin.cox@jila.colorado.edu}

\date{\today}

\begin{abstract}
We demonstrate that reversible dephasing via inhomogeneous broadening can greatly reduce collective quantum state rotation errors, and observe the suppression of rotation errors by more than 21~dB in the context of collective population measurements of the spin states of an ensemble of $2.1 \times 10^5$ laser cooled and trapped $^{87}$Rb atoms.  The large reduction in rotation noise enables direct resolution of spin state populations 13(1) dB below the fundamental quantum projection noise limit.  Further, the spin state measurement projects the system into an entangled state with 9.5(5) dB of directly observed spectroscopic enhancement (squeezing) relative to the standard quantum limit, whereas no enhancement would have been obtained without the suppression of rotation errors.
\end{abstract}


\maketitle 

Decoherence destroys entanglement, degrades precision measurement signals, and limits a wide range of coherent processes from lasing to operating quantum gates \cite{Townes_Laser_1958, Ladd_QuantumComputer_2010}.  Therefore, most technologies relying on real or synthetic atoms try to minimize decoherence resulting from loss, relaxation, and inhomogeneous broadening.  Recently, however, specifically engineered forms of decoherence have been used to enhance certain processes.  Dissipative decoherence, for example, can remove information from a system leading to stabilization of polar molecules from lossy collisions \cite{YMG13} or generation of entanglement \cite{FDT12, KMJ11,RTJ13}.  Also, non-dissipative, reversible, decoherence in the form of inhomogeneous broadening can be used to stabilize coherent operations allowing, for example, storage of non-classical light signals \cite{HSC11}, or as we show in this Letter, insensitivity to errors in collective quantum state rotations.

Precision measurements using one or many atoms require precise rotations of the atoms' quantum state.  These rotations, achieved by applying a coherent field at or near the atomic transition frequency, are used to excite an atomic transition \cite{Rabi_1939, BloomNature2014}, map the evolution of a quantum phase into a measurable quantity \cite{Ramsey_separatedfields_1950, Hinkley13092013}, or simply transfer state populations for precision readout \cite{CBS11}.  Imperfections in these rotations lead to classical uncertainty in the atoms' quantum state, which can dominate fundamental quantum uncertainty and limit precision measurements.  

In this Letter, we propose and experimentally demonstrate an approach to suppress rotation errors using reversible inhomogenous broadening, an alternative to the composite coupling pulses that are often used to correct state rotation errors \cite{Tycho_composite_1985,Wimperis_composite_1994,Khaneja_Composite_2005,Li_Composite_2006,PulseSequence_Uhrig_2007,Steffen_Composite_2007,Rakreungdet_Composite_2009,Torosov_Composite_2011}.  We first theoretically show how collective rotations of many qubits can be performed with greatly reduced errors if controlled inhomogeneous broadening of the transition is applied prior to the desired rotation.  We also show that collective coherence is restored by reversal of the inhomogeneous broadening after the rotation.  

\begin{figure}[t!]
\includegraphics{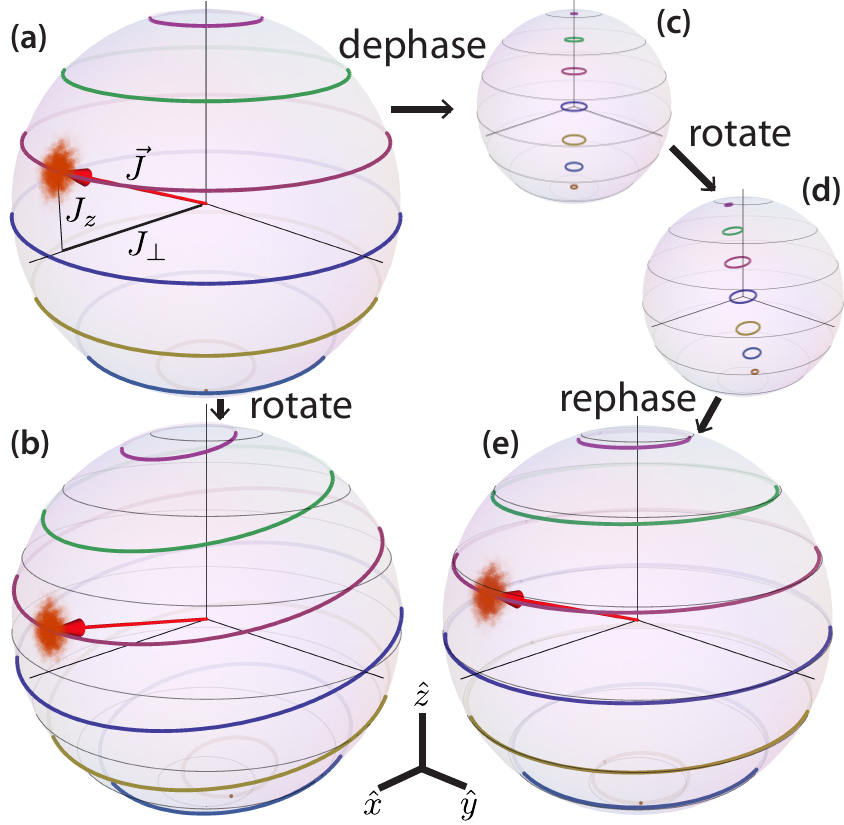}
\caption{\label{fig:fig1}
The reduced sensitivity of a dephased rotation to a small rotation error is graphically represented on collective Bloch spheres. A single representative Bloch vector prepared in the $\hat{x}-\hat{z}$ plane, along with the quantum uncertainty in its position, is shown (red arrow and noise distribution).  Each sphere also has a series of colored lines denoting the tips of Bloch vectors that are at a constant $J_z$ in the initial configuration.  The original rings of constant $J_z$ are shown in parts (b-e) as thin black lines for reference.  A small rotation $\mathcal{R}_{\hat{y}}(\pi/16)$ representing an error is applied without (b) and with (c,d,e) dephasing.  By reversibly dephasing the Bloch vector to $C_d = 0.14$, the impact of the rotation is greatly reduced.  Rotation errors that would otherwise dominate can be supressed well below the fundamental quantum noise.}
\end{figure}

Next we apply dephased rotations in a specific experiment, demonstrating a maximum suppression of technical noise of greater than $21$~dB when rotating the internal states of laser-cooled and trapped $^{87}$Rb atoms.  Dephased rotations aid the generation and observation of entangled, spin-squeezed states with a directly observed enhancement in quantum phase estimation $9.5(5)$~dB below the standard quantum limit for an unentangled ensemble, one of the largest such enhancements in atomic systems reported to date \cite{Squeezing_Bohnet_2013, PhysRevLett.112.155304, ChapmanSqueeze,BlattSqueeze,LSM10}. In the absence of any reversible inhomogeneous broadening, either incidental or deliberate, we estimate that little to no squeezing would have been observed in this experiment due to imperfections in the required quantum state rotations \footnote{A different measurement sequence (Ref. \citenum{Squeezing_Bohnet_2013}) that avoids any rotations achieved a comparable amount of directly observed spin squeezing in the same system without relying on dephased rotations, but only by sacrificing a factor of two in fundamental measurement resolution.  This additional factor of two was not realized in this work due to the probe laser's frequency noise coupling more strongly into the measurment sequence of Fig.~\ref{fig:fig3} }.

Dephased rotations are a general concept and could be applied to a variety of applications, having several advantages over traditional composite rotation sequences.  Composite sequences rely on cancellation between the errors of each individual rotation. However, cancellation fails if the errors fluctuate on time scales comparable to the time required for the composite pulse sequence.  Furthermore, increasing the rate of rotations to enhance the correlation in errors may actually be detrimental depending on the form of the noise spectrum \cite{PhaseNoise_Chen_2012}.  Lastly, composite pulses require precise control over the phase of the coupling field, and the most effective composite sequences require many pulses, increasing the time required for a measurement sequence.  The approach presented here to a large degree avoids these requirements.  We note that intense efforts to apply composite pulses to reduce rotation-added noise were largely unsuccessful in our experiment.

We describe our system of $N$ 2-level atoms as spin-1/2 particles using a collective Bloch vector ${\bf J} = J_x \hat{x}+J_y \hat{y}+J_z \hat{z}= \Sigma_{i=1}^N {\bf J}_i $, where the $i$th Bloch vector ${\bf J}_i = \langle \hat{{\bf J}}_i \rangle $ is the expectation value of the quantum spin projection operator for the $i$th atom. The $\hat{z}$ projection of the collective Bloch vector $J_z \equiv {\bf J} \cdot \hat{z} = \left(N_{\uparrow}-N_{\downarrow}\right)/2$, is directly determined by measuring the number of atoms in spin up $N_\uparrow$ and down $N_\downarrow$.  Precision measurements with 2-level systems are fundamentally limited by quantum uncertainty in the angles describing the orientation of the Bloch vector. This quantum uncertainty appears as quantum projection noise (QPN) in the measurement of the spin projection $J_z$.   For unentangled atoms,  the rms fluctuation for a coherent spin state (CSS) with ${\bf J}= N/2~\hat{x}$ is $\Delta J_{z,QPN}=\sqrt{N}/2$.  The projection noise limits the estimate of the Bloch vector's polar angle to an rms uncertainty of $\Delta \theta_{SQL} = 1/\sqrt{N}$, the so-called standard quantum limit (SQL). Due to this scaling, states with large $N$ are desirable for precise phase estimation, but in these states, classical rotation errors become more challenging to reduce below the smaller SQL.  

The rotation of the $i$th Bloch vector through angle $\psi_i$ about an axis $\hat{n}$ is defined by the rotation matrix $\mathcal{R}_{\hat{n}}(\psi_i)$.  If the rotation is uniform ($\psi_i = \psi$ for all $i$) then the result is a rigid rotation in which the length of the Bloch vector is conserved. The errors we wish to suppress are those generated by uniform rotation errors associated with the coupling field, in particular, an arbitrary erroneous rotation through a small angle $\phi$ described by $\mathcal{R}_{\hat{n}}(\phi)$.  The suppression of the rotation errors will be achieved by introducing a brief, controlled inhomogeneous broadening of the energy difference between $\ket{\uparrow}$ and $\ket{\downarrow}$ before and after the imperfect rotation.  The time-integrated effect of the broadening on the $i$th vector is characterized by the non-uniform rotation $\mathcal{R}_{\hat{z}}(\psi_i)$. The amount of dephasing is quantified by the fractional reduction in the collective Bloch vector's transverse projection $J_\perp\equiv \sqrt{J_x^2+J_y^2}$.  Specifically, we define the transverse coherence $C_d = J_{\perp d}/J_{\perp 0}$, where the subscript $d$ refers to $J_\perp$ after dephasing and $0$ refers to $J_\perp$ prior to dephasing.  In the present work, the inhomogeneous broadening will be achieved through light shifts, but could also be realized through magnetic fields or electric fields.  Whatever method is used, the key is that the dephasing must be reversible:  at a later time the opposite rotation can be realized $\mathcal{R}_{\hat{z}}(-\psi_i)$ to fully or partially undo the dephasing.  Here the dephasing will be undone by using a $\pi$-pulse (e.g. $\mathcal{R}_{\hat{y}}(\pi)$) followed by identical inhomogeneous broadening.

To theoretically show that dephased collective spin vectors are protected from small rotation errors, we analyze the rotation error of a nominal $\pi$-pulse, with fractional amplitude error $\epsilon$ and detuning error $\delta$ of the applied coupling field from the atomic transition, that is preceded and followed by dephasing steps. The final Bloch vector after such a sequence is 
\begin{equation}
{\bf J}_F=\Sigma_{i=1}^N \mathcal{R}_{\hat{z}}(\psi_i)\mathcal{R}_{\hat{\gamma}}(\beta)\mathcal{R}_{\hat{z}}(\psi_i){\bf J}_{i0} \,,
\end{equation}
\noindent where the subscript $F$ indicates a quantity after all rotations.

The effective rotation angle is a function of both $\epsilon$ and $\delta$ and can be written $\beta = \pi \sqrt{(1+\epsilon)^2+\delta^{*2}}$ where $\delta^* = \delta/ \Omega$ and $\Omega$ is the on resonance Rabi frequency of the applied rotation.  In the rotating frame of the applied field, the rotation axis depends on the detuning error, $\hat{\gamma}\propto \Omega \hat{\alpha}+\delta \hat{z}$.  For an arbitrary initial Bloch vector, the rotation axis $\hat{\alpha}=\hat{y}$ can be chosen without loss of generality.

As an example, we assume that the inhomogeneous phase rotation angles $\psi_i$ are drawn from a Gaussian distribution with mean of zero and rms value $\sigma$.  The reduction in transverse coherence due to the applied inhomogeneous broadening in this case is $C_d = e^{-\sigma^2/2}$.  The complete sequence of applied broadening and imperfect rotations can then be averaged over all atoms to compute the final Bloch vector ${\bf J}_F$ with solution,

\begin{equation}
\label{theoryresults}
\left(
\begin{array}{c}
J_{xF}\\
J_{yF}\\
J_{zF}\\
\end{array}
\right)
\approx -
\left(
\begin{array}{l}
J_{x0}(1-\eta^2)+C_d \pi\epsilon  J_{z0} \\
J_{y0}(1-\eta^2) - C_d 2\delta^* J_{z0} \\
J_{z0}(1-2\eta^2) - C_d ( 2\delta^*J_{y0} + \pi\epsilon J_{x0}). \\
\end{array}\right)
\end{equation}

\noindent where $\eta^2\equiv \pi^2\epsilon^2/4+\delta^{*2}$.  We have assumed here that $\pi \epsilon$, $\delta^*$, and $C_d$ $\ll 1$, and neglected all terms of third order in products of these quantities.  

The key result is that all rotation errors that are first order in $\pi \epsilon$ and $\delta^*$ are reduced by a factor $C_d$.  The cost of this error suppression is shortening of the Bloch vector, but only at second order in the rotation error $\eta$.  The final transverse Bloch vector component $C_F = J_{\perp F}/J_{\perp 0}$ is reduced as $C_F\approx 1-\eta^2$, and the $\hat{z}$ projection of the Bloch vector is reduced to $J_{zF}/J_{z0}\approx1-2\eta^2$.  

Fig. 1 graphically demonstrates the reduced sensitivity of an arbitrary CSS to a rotation about an axis on the equator.  The rings of constant color indicate the location of the tips of the Bloch vectors with equal $J_z$ at the beginning of a rotation sequence (top left).  Subsequent steps indicate how these points are mapped to new positions due to rotations and dephasing, with the initial ring locations shown in black for reference.  The figure depicts the effect of an error $\pi\epsilon=\pi/16$, $\delta^* = 0$ rotation about the $y$-axis with and without dephasing to $C_d = 0.14$, a reasonable experimental value.  Without dephasing, the rotation error can cause angular deflections greater than the representative quantum noise distribution (shown for $N = 120$ for visual clarity).  With dephasing, the rotation error is greatly reduced, causing negligible error compared to the quantum noise. 

The dephased rotation scheme exhibits an additional useful attribute for suppressing rotation errors in the generation and manipulation of spin-squeezed ensembles.  Dephased rotations can significantly reduce the amount of anti-squeezing projected into the low noise squeezed quadrature by a rotation error.  We show this theoretically in the Supplementary Material.

\begin{figure}[!ht]
\includegraphics{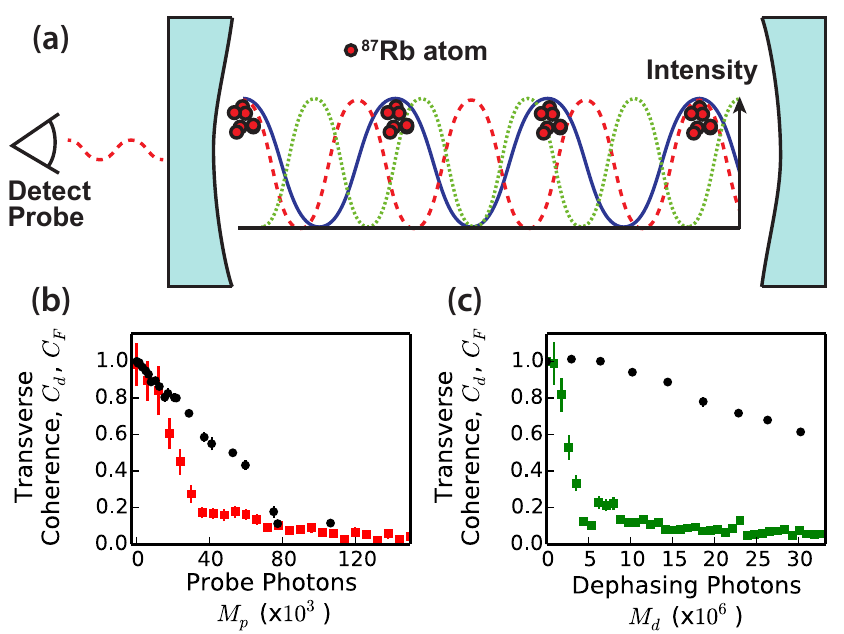}
\caption{\label{fig:fig2}(a) The standing wave intensity of each beam is shown inside the cavity (blue mirrors).  The atoms are trapped at antinodes of the 823~nm optical lattice (blue).  The probe laser at 780~nm (red dashed) and dephasing beam at 795~nm (green dotted) cause dephasing due to their inhomogeneous light shifts.  We detect the phase of the probe light to infer $N_\uparrow$.  (b,c) The reduction in transverse coherence after dephasing $C_d$ (red and green squares) and rephasing $C_F$ (black circles) is measured versus the average number of transmitted photons from the probe beam (b) and dephasing beam (c).}
\end{figure}
 
We apply the proposed scheme to collective measurements of $N=2.1\times10^5$ to $N=5\times10^5$~$^{87}$Rb atoms laser-cooled and trapped inside an optical cavity of finesse $F=660$ (see ref. \citenum{Squeezing_Bohnet_2013} for experimental details).  The atoms are tightly confined by a 1D optical lattice formed by exciting a longitudinal TEM$_{00}$ mode of the cavity with wavelength $\lambda_l= 823$~nm.  The atoms fill lattice sites long the central 2~mm of the cavity.  The spin system is defined by hyperfine ground states $\ket{\uparrow}= |F=2,m_f=2\rangle$ and $\ket{\downarrow}= |F=1,m_f=1\rangle$.  Coherent rotations between these states are performed by applying microwaves at the transition frequency $6.83$~GHz.  $N_\uparrow$ can be inferred by measuring the dispersive frequency shift of another TEM$_{00}$ cavity mode tuned $\approx 200$~MHz from resonance with the optical transition between $ \ket{\uparrow} $ and an excited state $\ket{e}= |F'=3,m_f=3\rangle$ on the 780~nm D2 line \cite{Squeezing_Bohnet_2013}.

The probe light at $\lambda_p=780$~nm that is used to measure the cavity frequency shift and infer $N_\uparrow$ also creates an inhomogeneous light shift that dephases the atoms.  Since the standing waves of the lattice and probe are incommensurate ($\lambda_p \neq \lambda_l$), the atoms at different lattice sites experience different light shifts from the probe, leading to dephasing (shown in Fig. \ref{fig:fig2}). 

We can also apply an additional dephasing laser tuned to resonance with yet another TEM$_{00}$ longitudinal mode of the cavity.  This dephasing beam is detuned $\approx50$~GHz from the 795~nm D1 optical transition and allows us to modify the amount of dephasing without modifying the signal to noise of the atom number probe or causing additional unwanted free-space scattering.  The 795~nm beam also serves to dephase the sub-class of atoms at lattice sites that are at anti-nodes of the probe mode (see Fig. \ref{fig:fig2}(a)).  Because the atoms are tightly confined with respect to the cavity axis, the same light shifts can be applied at a later time. after a $\pi$-pulse, to reverse the applied phase shifts.

We can measure $C_d$ due to dephasing from the probe and dephasing lasers by first preparing a coherent spin state along $\hat{x}$.  We then apply either the probe or dephasing laser for a varying amount of time, after which we apply the rotation  $R_{\hat{\alpha}}(\pi/2)$ about a random axis $\hat{\alpha}$ lying in the $\hat{x}$-$\hat{y}$ plane. Lastly, we measure the number of atoms $N_\uparrow$.  When averaged over all rotation axes, the standard deviation of $N_\uparrow$ is proportional to $C_d$.  In Fig.~\ref{fig:fig2}(b) and (c), $C_d$ and $C_F$ are plotted versus the average number of probe $M_p$ and dephasing $M_d$ photons transmitted through the cavity.  For small $M_d$ the transverse coherence only shows second order reduction ($1-C_F \propto M_d^2)$), due to the large detuning of the dephasing beam from the optical transition.  In contrast, for small $M_p$ the transverse coherence loss is linear ($1-C_F \propto M_p$) due to the higher probability of single-atom wave function collapse from free-space scattering of probe photons.

In Fig. \ref{fig:fig3}(a) we demonstrate reduced sensitivity to rotation noise arising from environmental noise sources using our dephased rotation scheme.  Data showing reduced sensitivity to intentionally applied rotation errors can be found in the Supplementary Material.  In our experiment, undesirable environmental rotation noise arises primarily from microwave amplitude noise and frequency fluctuations in the magnetic field-sensitive hyperfine transition.  To demonstrate a reduction in sensitivity to environmental noise sources, $J_{zF}$ is measured after a large even number of $\pi$-pulses.  With increased dephasing, the rotation-added noise can be reduced below QPN even after eight $\pi$-pulses.

We now show how dephased rotations can be used in experiments to generate entangled, spin squeezed states by making precise collective measurements of the spin projection $J_z$.  These experiments are treated in detail in a related work \cite{Squeezing_Bohnet_2013}.  Here we primarily emphasize the role dephased rotations can play, enabling large reductions in technical rotation noise and allowing resolution of the spin projection far below the quantum projection noise level.  In our experiment, dephased rotations are highly advantageous to composite pulse sequences because they do not require any control of the applied rotation axis and do not increase the duration of the measurement sequence, which would increase sensitivity to low frequency noise.

To verify that the noise in $J_z$ is below QPN, two consecutive measurements of $J_z$, labeled $J_{z,p}$ and $J_{z,f}$ must be correlated below $\Delta J_{z,QPN}$.  
\begin{figure}
\includegraphics{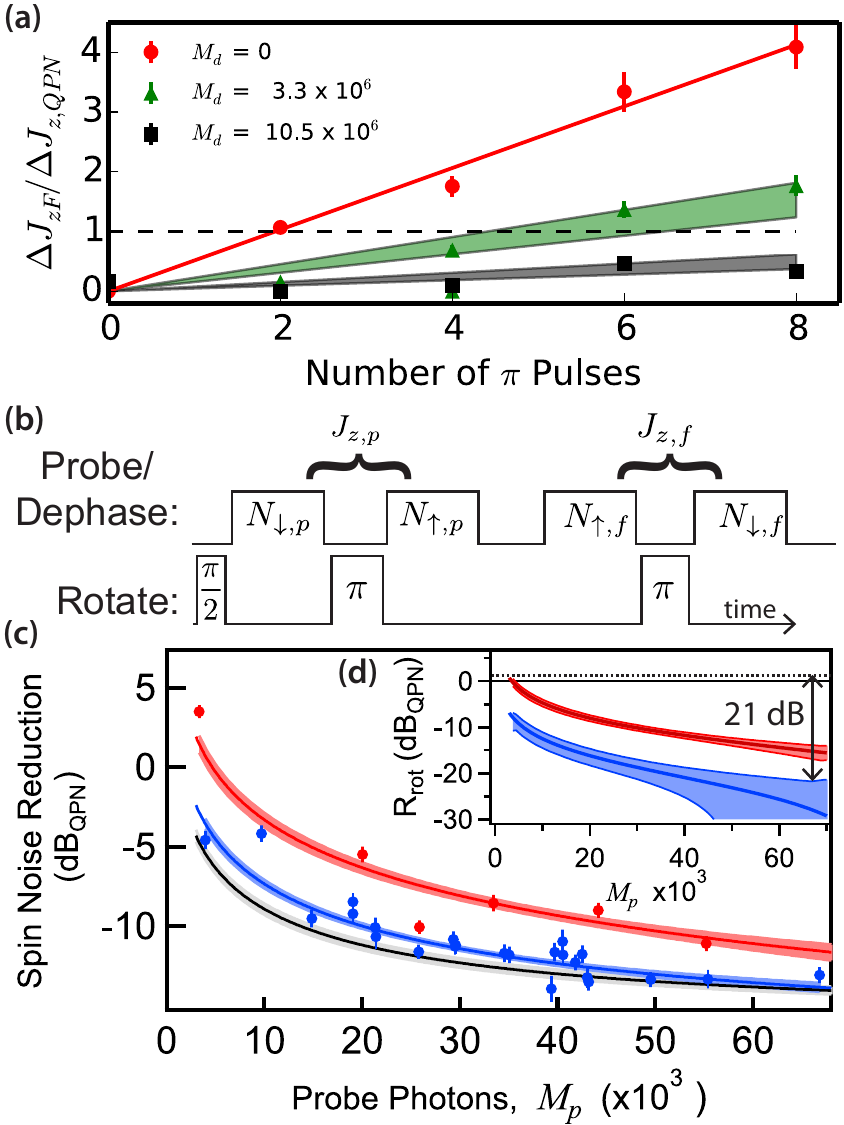}
\caption{\label{fig:fig3}(a) The rms noise in the measured spin projection $J_{zF}$, $\Delta J_{zF}$,  after applying an integer number of $\pi$-pulses is displayed for three different amounts of applied dephasing, quantified by $M_d$.  The contribution to $\Delta J_{zF}$ due to finite measurement resolution (i.e. $\Delta J_{zF}$ at 0 $\pi$-pulses) is subtracted out.  For $M_d = 0$, a linear fit extracts the rotation-added noise per pulse (red line).  Predictions (green and black bands) using $C_d$ from Fig.~\ref{fig:fig2} reasonably explain the reduction in rotation-added noise with increased $M_d$.  All shaded regions represent 68\% confidence intervals.  (b) Dephased rotations are applied in a sequence designed to resolve spin populations below QPN.  $N_\uparrow$ is measured before and after a $\pi$-pulse with outcomes labeled $N_{\uparrow,p}$, $N_{\downarrow,p}$, $N_{\uparrow,f}$ and $N_{\downarrow,f}$ to determine the spin noise reduction $R$.  Both the 780~nm probe and 795~nm dephasing beams are applied during each measurement of $N_\uparrow$ and $N_\downarrow$.  (c) $R$ is measured as a function of probe strength $M_p$ for $M_d = 6.1(3) \times 10^6$ (blue data and fit) and $M_d = 0$ (red data and fit).  All quantities are displayed in units of dB relative to QPN, dB$_{\textrm{QPN}}$.  The fit to the measurement background $R_{\textrm{bck}}$ is shown in black.  (d) The rotation-added noise $R_{\textrm{rot}}$, shown in the inset, can be inferred from the data of part (c).  $R_{\textrm{rot}}$ with no dephasing is shown as a dashed line.  Dephasing can reduce $R_{\textrm{rot}}$ by more than $21$~dB. }
\end{figure}
The degree of spin noise compared to quantum projection noise is characterized by the spin noise reduction $R = [\Delta(J_{z,f}-J_{z,p})]^2/\Delta J_{z,QPN}^2$, where $\Delta(J_{z,f}-J_{z,p})$ is the standard deviation in the differential quantity $J_{z,f}-J_{z,p}$.  The measurement sequence for $R$ is shown in Fig. \ref{fig:fig3}(b).  The spin noise reduction has two contributions $R = R_{\textrm{bck}}+R_{\textrm{rot}}$.  One, $R_{\textrm{bck}}$, we attribute to measurement imprecision of the experiment along with measurement back-action.  The other, $R_{\textrm{rot}}$, is rotation-added noise from the two $\pi$-pulses in the measurement sequence.  We estimate $R_{\textrm{bck}}$ (black line in Fig. \ref{fig:fig3}(c)) by performing the measurement sequence of \ref{fig:fig3}(b) without the $\pi$-pulses.  The experiment is then repeated with the $\pi$-pulses included, and any increase in $R$ is assigned as rotation-added noise $R_{\textrm{rot}}$.

In Fig.~\ref{fig:fig3}(c), the measured spin noise reduction and measurement background  are shown versus $M_p$ (transmitted probe photons in a single measurement window).  With the probe beam alone (i.e.~$M_d=0$), the spin noise reduction $R$ (red data and fit)  lies well above the measurement background  $R_{\textrm{bck}}$ (black line).  However, when the additional dephasing is applied with strength $M_d = 6.1(3)\times 10^6$, the observed $R$ (blue points and fit) is improved to values very close to the measurement background.

Fig. \ref{fig:fig3}(d) (inset) displays the inferred rotation-added noise $R_{\textrm{rot}}$ with and without the additional dephasing applied (blue and red lines respectively).  The combined dephasing of the probe and dephasing beams allows a reduction of the rotation-added noise of greater than approximately $21$~dB compared to the original rotation noise with no dephasing  (i.e. $R_{\textrm{rot}}\approx 0$~dB$_{\textrm{QPN}}$ when $M_p=0$ and $M_d=0$) enabling up to $R=13(1)$~dB of spin noise reduction below the QPN at $2.1\times 10^5$ atoms.

The rephasing nearly completely restores coherence, as demonstrated in Fig. \ref{fig:fig2}.  As a result, the state generated after the premeasurement can be viewed as a determinstically generated spin-squeezed state (i.e. no post-selection), conditioned on knowledge of the measurement outcome $J_{z,p}$ on a given trial.  After accounting for both the degree of spin-noise reduction $R$ and the loss of coherence $C_F$, the optimum measurement sequence with dephasing provides a directly observed enhanced phase resolution $9.5(5)$ dB below the SQL.  In contrast, without any reversible dephasing, rotation-added noise would have precluded the observation of any enhancement beyond the SQL.

\begin{acknowledgments}
The authors would like to acknowledge helpful discussions 
with Zilong Chen. All authors acknowledge financial support from DARPA QuASAR,
ARO, NSF PFC, and NIST.  K.C.C. acknowledges support
from NDSEG. This work is supported by the National
Science Foundation under Grant Number 1125844.
\end{acknowledgments}

\bibliography{main}

\pagebreak
\widetext
\begin{center}
\textbf{\large Reducing Collective Quantum State Rotation Errors with Reversible Dephasing:  Supplementary Material}
\end{center}
\setcounter{equation}{0}
\setcounter{figure}{0}
\setcounter{table}{0}
\setcounter{page}{1}
\makeatletter
\renewcommand{\theequation}{S\arabic{equation}}
\renewcommand{\thefigure}{S\arabic{figure}}

\section{Reduction of Applied Frequency and Amplitude Errors}

The main text shows both experimentally and theoretically that dephasing reduces an ensemble's sensitivity to collective rotation errors arising from imperfections in the coupling field used for state manipulation.    
To gain intuition, consider the case when the initial Bloch vector lies on the equator of the Bloch sphere ($J_z = 0$).  In this case, the length of the dephased Bloch vector $J_d$ is reduced, by definition of $C_d$, to $J_d = C_d J_0$.  Since the length of the Bloch vector is reduced by $C_d$, the possible change in the Bloch vector's $J_z$ projection due to a rotation must also be reduced by $C_d$.  

\begin{figure}
\includegraphics{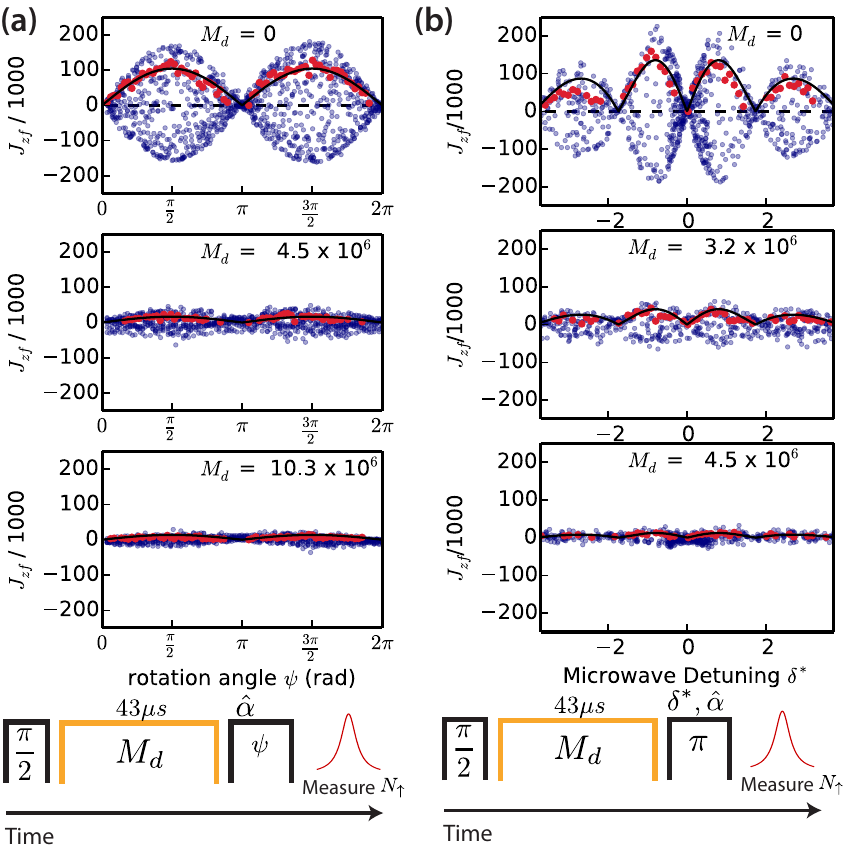}
\caption{\label{figS1}Reduction in sensitivity to amplitude and frequency rotation errors.  (a) The ensemble is subjected to a rotation with arbitrary amplitude $\psi$ and equitorial rotation axis $\hat{\alpha}$ and zero detuning (measurement sequence shown below graph).  The resulting $J_{zf}$ (blue points) are plotted versus amplitude of the rotation $\psi$ for three different values of dephasing, characterized by $M_d$.  At each amplitude the average magnitude of $J_{zf}$ (red points) is compared to a prediction using the measured transverse coherence from Fig. 2 (black line).  (b)  The ensemble is subjected to a rotation with aribitrary detuning $\delta^*$ and azimuthal rotation axis $\hat{\alpha}$.  The amplitude is constrained so that at zero detuning, the rotation is a $\pi$-pulse.  $J_{zf}$ (blue points) are plotted as a function of $\delta^*$, again for three different values of dephasing, and the average magnitude of $J_{zf}$ (red points) is in agreement to a prediction from the measured transverse coherence (black line).}
\end{figure}

Fig. \ref{figS1} demonstrates this reduced sensitivity to intentionally applied rotations representing amplitude and frequency errors in the coupling field.  Measurement sequences are shown at the bottom of Fig. \ref{figS1}.  A $\pi/2$-pulse initializes the Bloch vector at the equator (black pulse).  Dephasing is applied with a strength characterized by the average number of photons transmitted through the cavity $M_d$ (yellow pulse).  A rotation representing an error (either amplitude (a) or detuning (b)) is applied, after which $N_\uparrow$ is measured by measuring the shift of the optical cavity resonance frequency\cite{Squeezing_Bohnet_2013,PhysRevA.89.043837}.  From the measured $N_\uparrow$, we infer the $z$-projection of the final Bloch vector $J_{zf}$. 

For part (a), the applied rotation is $\mathcal{R}_{\hat{\alpha}}(\psi)$, where $\psi$ is the arbitrary rotation amplitude, and $\hat{\alpha}$ is a random rotation axis lying in the $\hat{x}-\hat{y}$ plane of the Bloch sphere.  $J_{zf}$ (blue points) is plotted as a function of the applied amplitude $\psi$ for three different values of dephasing.   The randomization of $\hat{\alpha}$ causes large scatter of $J_{zf}$ over positive and negative values.  To compare with an expectation, we plot the average magnitude of the measured $J_{zf}$ in red, and a prediction based on the independently measured transverse coherence $C_d$ from Fig. 2 of the main text is shown as a black line.  The envelope of the data decays linearly with $C_d$ in reasonable agreement Eq. 2 in the main text and our intuitive expectation.
 

To demonstrate the reduction in sensitivity to rotations for which the coupling field is detuned from the atomic resonance frequency, we apply a nominal $\pi$-rotation with variable detuning.  The applied rotation is $\mathcal{R}_{\hat{\gamma}}(\pi\sqrt{1+\delta^{*2}})$ where $\hat{\gamma} \propto \Omega \hat{\alpha} + \delta \hat{z}$, and the azimuthal axis $\hat{\alpha}$ is randomized between each trial.  $\delta^* = \frac{\delta}{\kappa / 2}$ is the detuning of the coupling field from atomic resonance in cavity half-widths.  In Fig. \ref{figS1}(b), $J_{zf}$ (blue points) is plotted versus $\delta^*$ for three different values of $M_d$, and the average magnitude of $J_{zf}$ (red points) are compared to a prediction (black line).  Just as with the amplitude errors, the magnitude of the deflections of $J_{zf}$ scale linearly  with $C_d$ in good agreement with the prediction.

\section{Dephased Rotation of Quantum Noise}

When dealing with spin squeezed states, classical rotation errors can rotate the anti-squeezed quadrature of a system into the measurement basis.  This leads to additional noise above the noise in classical rotation errors primarily considered in the main text.  In this section, we treat this problem theoretically with a fully quantum mechanical description of the spin state, instead of treating each spin as a classical vector as in the main text.  We show that dephasing protects squeezed noise distributions from rotation of the anti-squeezed spin projection into the originally squeezed quadrature.  

We assume for simplicity that there is an arbitrary initial state oriented along $\hat{y} $ with a squeezed noise distribution in $\hat{J}_z$ (and anti-squeezed in $\hat{J}_x$).  The state is subjected to a small rotation amplitude error of size $\pi \epsilon$ around the $\hat{y}$-axis which rotates the anti-squeezed spin projection into the $z$ quadrature.  The initial squeezed noise distribution is characterized by the second order expectation value


\begin{align}
\langle \hat{J}_{z}^2\rangle_0 \equiv \bra{\psi_0}\hat{J}_z^2\ket{\psi_0} ,
\end{align}
where $\ket{\psi_0}$ describes the initial state. We wish to evaluate the noise distribution of $\hat{J}_z$ for the final state,
\begin{align}
\label{eq2}
\langle  \hat{J}_{z}^2 \rangle_f \equiv\bra{\psi_f}\hat{J}_z^2\ket{\psi_f},
\end{align}
where the final state $\ket{\psi_f}$ can be written in terms of $\ket{\psi_0}$ using the dephasing operator
\begin{align}
\hat{D}\equiv \prod_{i}^N \hat{R}_{\hat{z}} (\theta_i)
\end{align}
and rotation operator (around $\hat{y}$) $\hat{R}_{\hat{y}}(\pi \epsilon)$,  $\ket{\psi_f}=\hat{D}^\dag \hat{R}_{\hat{y}}(\pi \epsilon) \hat{D} \ket{\psi_0}$.  For now, we do not need to specify a specific form of dephasing (i.e. the inhomogeneous rotations $\theta_i$).  Using the dephasing and rotation operators, Eq. \ref{eq2} becomes

\begin{align}
\label{algebra}
\langle  \hat{J}_{z}^2 \rangle_f= \bra{\psi_0}D^\dag \hat{R}_{\hat{y}}(\pi \epsilon)^\dag \hat{D}  \hat{J}_z^2 \hat{D}^\dag  \hat{R}_{\hat{y}}(\pi \epsilon) \hat{D} \ket{\psi_0}.
\end{align}
The rotation operators can be written in terms of single atom spin projection operators as, to second order in the small parameter $\pi \epsilon$, 
\begin{align}
\hat{R}_{\hat{y}}(\pi \epsilon) \approx \prod_{k}^N (\mathbb{I}_k+i \pi \epsilon \hat{J}_{\hat{y},k}-\frac{(\pi \epsilon)^2}{2}\mathbb{I}_k),
\end{align}
where $\mathbb{I}_k$ is the identity operator for the $k$th atom, and the collective spin projection operator $\hat{J}_{\hat{y}}$ can be written as a sum over all atoms' individual spin operators $\hat{J}_{\hat{y}}=\sum_{i}^N \hat{\textbf{J}}_{i}\cdot \hat{y}$.  Using these definitions and single atom commutation relations, we simplify Eq.~\ref{algebra} keeping to second order in $\epsilon$,

\begin{align} 
\label{result}
\langle \hat{J}_z^2 \rangle_f = \langle \hat{J}_z^2 \rangle_0+ (\pi \epsilon)^2  \langle \hat{J}_{x}^2 \rangle_{d} -(\pi \epsilon)^2  \langle   \hat{J}_{z}^2 \rangle_0 + 2 \pi \epsilon \langle  \hat{J}_{z} \hat{J}_{x} \rangle_d.
\end{align}
The expectation values in the second and final terms (with subscript $d$) are calculated with respect to the dephased state, $\ket{\psi_d}= \hat{D} \ket{\psi_0}$.  This equation is particularly useful because it gives the quantum noise rotation in terms of measureable quantities for an arbitrary form of the dephasing.

For squeezed states oriented along $\hat{y}$ with symmetry around $\hat{x}$ and $\hat{z}$ (generated, for example, by 2-axis twisting or quantum non-demolition measurement \cite{Squeezing_Bohnet_2013,kitagawa_1993}) the final term in Eq. \ref{result} is zero, giving
\begin{align} 
\label{resultsimp}
\langle \hat{J}_z^2 \rangle_f = \langle \hat{J}_z^2 \rangle_0+(\pi \epsilon)^2  \langle \hat{J}_{x}^2 \rangle_d -  (\pi \epsilon)^2\langle   \hat{J}_{z}^2 \rangle_0.
\end{align}
This result shows that the back-action quadrature is introduced at order $\epsilon^2$ through  $\langle \hat{J}_{x}^2 \rangle_d$ instead of $\langle \hat{J}_{x}^2 \rangle_0$.  
In the limit of random Gaussian dephasing (as considered in the main text), $\langle \hat{J}_{x}^2 \rangle_d$ can be written,
\begin{align}
\label{eqJd}
\langle \hat{J}_{x}^2 \rangle_d =  C_d^2 \langle \hat{J}_{x}^2 \rangle_0 + (1-C_d^2) \Delta J_{QPN}^2.
\end{align}
In the limit of small and moderate dephasing, the standard deviation of the back-action quadrature that is rotated into $\hat{z}$ is reduced linearly with $C_d$ (first term).  However, for complete random dephasing, the back-action can only be reduced to the quantum projection noise level for a CSS, $\Delta J_{QPN} = \sqrt{N}/2$ as seen by the second term.  Applying greater dephasing provides marginal returns when the two terms in Eq. \ref{eqJd} become equal.  This occurs at a value of $C_d$ which we label $C_d'$, 
\begin{align}
C_d'^2=\frac{1}{1+\langle \hat{J}_{x}^2 \rangle_0 / \Delta J_{QPN}^2}.
\end{align}
For a spin squeezed state with a large back-action quadrature, $\langle \hat{J}_{x}^2 \rangle_0 \gg \Delta J_{QPN}^2$, more dephasing is required to reduce $\langle \hat{J}_{x}^2 \rangle_d$, the dephased back-action projection, to near the QPN level.

\bibliography{main.bib}

\begin{thebibliography}{29}%
\makeatletter
\providecommand \@ifxundefined [1]{%
 \@ifx{#1\undefined}
}%
\providecommand \@ifnum [1]{%
 \ifnum #1\expandafter \@firstoftwo
 \else \expandafter \@secondoftwo
 \fi
}%
\providecommand \@ifx [1]{%
 \ifx #1\expandafter \@firstoftwo
 \else \expandafter \@secondoftwo
 \fi
}%
\providecommand \natexlab [1]{#1}%
\providecommand \enquote  [1]{``#1''}%
\providecommand \bibnamefont  [1]{#1}%
\providecommand \bibfnamefont [1]{#1}%
\providecommand \citenamefont [1]{#1}%
\providecommand \href@noop [0]{\@secondoftwo}%
\providecommand \href [0]{\begingroup \@sanitize@url \@href}%
\providecommand \@href[1]{\@@startlink{#1}\@@href}%
\providecommand \@@href[1]{\endgroup#1\@@endlink}%
\providecommand \@sanitize@url [0]{\catcode `\\12\catcode `\$12\catcode
  `\&12\catcode `\#12\catcode `\^12\catcode `\_12\catcode `\%12\relax}%
\providecommand \@@startlink[1]{}%
\providecommand \@@endlink[0]{}%
\providecommand \url  [0]{\begingroup\@sanitize@url \@url }%
\providecommand \@url [1]{\endgroup\@href {#1}{\urlprefix }}%
\providecommand \urlprefix  [0]{URL }%
\providecommand \Eprint [0]{\href }%
\providecommand \doibase [0]{http://dx.doi.org/}%
\providecommand \selectlanguage [0]{\@gobble}%
\providecommand \bibinfo  [0]{\@secondoftwo}%
\providecommand \bibfield  [0]{\@secondoftwo}%
\providecommand \translation [1]{[#1]}%
\providecommand \BibitemOpen [0]{}%
\providecommand \bibitemStop [0]{}%
\providecommand \bibitemNoStop [0]{.\EOS\space}%
\providecommand \EOS [0]{\spacefactor3000\relax}%
\providecommand \BibitemShut  [1]{\csname bibitem#1\endcsname}%
\let\auto@bib@innerbib\@empty
\bibitem [{\citenamefont {Schawlow}\ and\ \citenamefont
  {Townes}(1958)}]{Townes_Laser_1958}%
  \BibitemOpen
  \bibfield  {author} {\bibinfo {author} {\bibfnamefont {A.~L.}\ \bibnamefont
  {Schawlow}}\ and\ \bibinfo {author} {\bibfnamefont {C.~H.}\ \bibnamefont
  {Townes}},\ }\href {\doibase 10.1103/PhysRev.112.1940} {\bibfield  {journal}
  {\bibinfo  {journal} {Phys. Rev.}\ }\textbf {\bibinfo {volume} {112}},\
  \bibinfo {pages} {1940} (\bibinfo {year} {1958})}\BibitemShut {NoStop}%
\bibitem [{\citenamefont {Ladd}\ \emph {et~al.}(2010)\citenamefont {Ladd},
  \citenamefont {Jelezko}, \citenamefont {Laflamme}, \citenamefont {Nakamura},
  \citenamefont {Monroe},\ and\ \citenamefont
  {O'Brien}}]{Ladd_QuantumComputer_2010}%
  \BibitemOpen
  \bibfield  {author} {\bibinfo {author} {\bibfnamefont {T.~D.}\ \bibnamefont
  {Ladd}}, \bibinfo {author} {\bibfnamefont {F.}~\bibnamefont {Jelezko}},
  \bibinfo {author} {\bibfnamefont {R.}~\bibnamefont {Laflamme}}, \bibinfo
  {author} {\bibfnamefont {Y.}~\bibnamefont {Nakamura}}, \bibinfo {author}
  {\bibfnamefont {C.}~\bibnamefont {Monroe}}, \ and\ \bibinfo {author}
  {\bibfnamefont {J.~L.}\ \bibnamefont {O'Brien}},\ }\href
  {http://dx.doi.org/10.1038/nature08812} {\bibfield  {journal} {\bibinfo
  {journal} {Nature}\ }\textbf {\bibinfo {volume} {464}},\ \bibinfo {pages}
  {45} (\bibinfo {year} {2010})}\BibitemShut {NoStop}%
\bibitem [{\citenamefont {Yan}\ \emph {et~al.}(2013)\citenamefont {Yan},
  \citenamefont {Moses}, \citenamefont {Gadway}, \citenamefont {Covey},
  \citenamefont {Hazzard}, \citenamefont {Rey}, \citenamefont {Jin},\ and\
  \citenamefont {Ye}}]{YMG13}%
  \BibitemOpen
  \bibfield  {author} {\bibinfo {author} {\bibfnamefont {B.}~\bibnamefont
  {Yan}}, \bibinfo {author} {\bibfnamefont {S.~A.}\ \bibnamefont {Moses}},
  \bibinfo {author} {\bibfnamefont {B.}~\bibnamefont {Gadway}}, \bibinfo
  {author} {\bibfnamefont {J.~P.}\ \bibnamefont {Covey}}, \bibinfo {author}
  {\bibfnamefont {K.~R.~A.}\ \bibnamefont {Hazzard}}, \bibinfo {author}
  {\bibfnamefont {A.~M.}\ \bibnamefont {Rey}}, \bibinfo {author} {\bibfnamefont
  {D.~S.}\ \bibnamefont {Jin}}, \ and\ \bibinfo {author} {\bibfnamefont
  {J.}~\bibnamefont {Ye}},\ }\href {\doibase 10.1038/nature12483} {\bibfield
  {journal} {\bibinfo  {journal} {Nature}\ }\textbf {\bibinfo {volume} {501}},\
  \bibinfo {pages} {521} (\bibinfo {year} {2013})}\BibitemShut {NoStop}%
\bibitem [{\citenamefont {Foss-Feig}\ \emph {et~al.}(2012)\citenamefont
  {Foss-Feig}, \citenamefont {Daley}, \citenamefont {Thompson},\ and\
  \citenamefont {Rey}}]{FDT12}%
  \BibitemOpen
  \bibfield  {author} {\bibinfo {author} {\bibfnamefont {M.}~\bibnamefont
  {Foss-Feig}}, \bibinfo {author} {\bibfnamefont {A.~J.}\ \bibnamefont
  {Daley}}, \bibinfo {author} {\bibfnamefont {J.~K.}\ \bibnamefont {Thompson}},
  \ and\ \bibinfo {author} {\bibfnamefont {A.~M.}\ \bibnamefont {Rey}},\ }\href
  {\doibase 10.1103/PhysRevLett.109.230501} {\bibfield  {journal} {\bibinfo
  {journal} {Phys. Rev. Lett.}\ }\textbf {\bibinfo {volume} {109}},\ \bibinfo
  {pages} {230501} (\bibinfo {year} {2012})}\BibitemShut {NoStop}%
\bibitem [{\citenamefont {Krauter}\ \emph {et~al.}(2011)\citenamefont
  {Krauter}, \citenamefont {Muschik}, \citenamefont {Jensen}, \citenamefont
  {Wasilewski}, \citenamefont {Petersen}, \citenamefont {Cirac},\ and\
  \citenamefont {Polzik}}]{KMJ11}%
  \BibitemOpen
  \bibfield  {author} {\bibinfo {author} {\bibfnamefont {H.}~\bibnamefont
  {Krauter}}, \bibinfo {author} {\bibfnamefont {C.~A.}\ \bibnamefont
  {Muschik}}, \bibinfo {author} {\bibfnamefont {K.}~\bibnamefont {Jensen}},
  \bibinfo {author} {\bibfnamefont {W.}~\bibnamefont {Wasilewski}}, \bibinfo
  {author} {\bibfnamefont {J.~M.}\ \bibnamefont {Petersen}}, \bibinfo {author}
  {\bibfnamefont {J.~I.}\ \bibnamefont {Cirac}}, \ and\ \bibinfo {author}
  {\bibfnamefont {E.~S.}\ \bibnamefont {Polzik}},\ }\href {\doibase
  10.1103/PhysRevLett.107.080503} {\bibfield  {journal} {\bibinfo  {journal}
  {Phys. Rev. Lett.}\ }\textbf {\bibinfo {volume} {107}},\ \bibinfo {pages}
  {080503} (\bibinfo {year} {2011})}\BibitemShut {NoStop}%
\bibitem [{\citenamefont {Reiter}\ \emph {et~al.}(2013)\citenamefont {Reiter},
  \citenamefont {Tornberg}, \citenamefont {Johansson},\ and\ \citenamefont
  {S\o{}rensen}}]{RTJ13}%
  \BibitemOpen
  \bibfield  {author} {\bibinfo {author} {\bibfnamefont {F.}~\bibnamefont
  {Reiter}}, \bibinfo {author} {\bibfnamefont {L.}~\bibnamefont {Tornberg}},
  \bibinfo {author} {\bibfnamefont {G.}~\bibnamefont {Johansson}}, \ and\
  \bibinfo {author} {\bibfnamefont {A.~S.}\ \bibnamefont {S\o{}rensen}},\
  }\href {\doibase 10.1103/PhysRevA.88.032317} {\bibfield  {journal} {\bibinfo
  {journal} {Phys. Rev. A}\ }\textbf {\bibinfo {volume} {88}},\ \bibinfo
  {pages} {032317} (\bibinfo {year} {2013})}\BibitemShut {NoStop}%
\bibitem [{\citenamefont {Hosseini}\ \emph {et~al.}(2011)\citenamefont
  {Hosseini}, \citenamefont {Sparkes}, \citenamefont {Campbell}, \citenamefont
  {Lam},\ and\ \citenamefont {Buchler}}]{HSC11}%
  \BibitemOpen
  \bibfield  {author} {\bibinfo {author} {\bibfnamefont {M.}~\bibnamefont
  {Hosseini}}, \bibinfo {author} {\bibfnamefont {B.}~\bibnamefont {Sparkes}},
  \bibinfo {author} {\bibfnamefont {G.}~\bibnamefont {Campbell}}, \bibinfo
  {author} {\bibfnamefont {P.}~\bibnamefont {Lam}}, \ and\ \bibinfo {author}
  {\bibfnamefont {B.}~\bibnamefont {Buchler}},\ }\href {\doibase
  10.1038/ncomms1175} {\bibfield  {journal} {\bibinfo  {journal} {Nat.
  Commun.}\ }\textbf {\bibinfo {volume} {2}},\ \bibinfo {pages} {174} (\bibinfo
  {year} {2011})}\BibitemShut {NoStop}%
\bibitem [{\citenamefont {Rabi}\ \emph {et~al.}(1939)\citenamefont {Rabi},
  \citenamefont {Millman}, \citenamefont {Kusch},\ and\ \citenamefont
  {Zacharias}}]{Rabi_1939}%
  \BibitemOpen
  \bibfield  {author} {\bibinfo {author} {\bibfnamefont {I.~I.}\ \bibnamefont
  {Rabi}}, \bibinfo {author} {\bibfnamefont {S.}~\bibnamefont {Millman}},
  \bibinfo {author} {\bibfnamefont {P.}~\bibnamefont {Kusch}}, \ and\ \bibinfo
  {author} {\bibfnamefont {J.~R.}\ \bibnamefont {Zacharias}},\ }\href {\doibase
  10.1103/PhysRev.55.526} {\bibfield  {journal} {\bibinfo  {journal} {Phys.
  Rev.}\ }\textbf {\bibinfo {volume} {55}},\ \bibinfo {pages} {526} (\bibinfo
  {year} {1939})}\BibitemShut {NoStop}%
\bibitem [{\citenamefont {Bloom}\ \emph {et~al.}(2014)\citenamefont {Bloom},
  \citenamefont {Nicholson}, \citenamefont {Williams}, \citenamefont
  {Campbell}, \citenamefont {Bishof}, \citenamefont {Zhang}, \citenamefont
  {Zhang}, \citenamefont {Bromley},\ and\ \citenamefont
  {Ye}}]{BloomNature2014}%
  \BibitemOpen
  \bibfield  {author} {\bibinfo {author} {\bibfnamefont {B.~J.}\ \bibnamefont
  {Bloom}}, \bibinfo {author} {\bibfnamefont {T.~L.}\ \bibnamefont
  {Nicholson}}, \bibinfo {author} {\bibfnamefont {J.~R.}\ \bibnamefont
  {Williams}}, \bibinfo {author} {\bibfnamefont {S.~L.}\ \bibnamefont
  {Campbell}}, \bibinfo {author} {\bibfnamefont {M.}~\bibnamefont {Bishof}},
  \bibinfo {author} {\bibfnamefont {X.}~\bibnamefont {Zhang}}, \bibinfo
  {author} {\bibfnamefont {W.}~\bibnamefont {Zhang}}, \bibinfo {author}
  {\bibfnamefont {S.~L.}\ \bibnamefont {Bromley}}, \ and\ \bibinfo {author}
  {\bibfnamefont {J.}~\bibnamefont {Ye}},\ }\href
  {http://dx.doi.org/10.1038/nature12941} {\bibfield  {journal} {\bibinfo
  {journal} {Nature}\ }\textbf {\bibinfo {volume} {506}},\ \bibinfo {pages}
  {71} (\bibinfo {year} {2014})}\BibitemShut {NoStop}%
\bibitem [{\citenamefont {Ramsey}(1950)}]{Ramsey_separatedfields_1950}%
  \BibitemOpen
  \bibfield  {author} {\bibinfo {author} {\bibfnamefont {N.~F.}\ \bibnamefont
  {Ramsey}},\ }\href {\doibase 10.1103/PhysRev.78.695} {\bibfield  {journal}
  {\bibinfo  {journal} {Phys. Rev.}\ }\textbf {\bibinfo {volume} {78}},\
  \bibinfo {pages} {695} (\bibinfo {year} {1950})}\BibitemShut {NoStop}%
\bibitem [{\citenamefont {Hinkley}\ \emph {et~al.}(2013)\citenamefont
  {Hinkley}, \citenamefont {Sherman}, \citenamefont {Phillips}, \citenamefont
  {Schioppo}, \citenamefont {Lemke}, \citenamefont {Beloy}, \citenamefont
  {Pizzocaro}, \citenamefont {Oates},\ and\ \citenamefont
  {Ludlow}}]{Hinkley13092013}%
  \BibitemOpen
  \bibfield  {author} {\bibinfo {author} {\bibfnamefont {N.}~\bibnamefont
  {Hinkley}}, \bibinfo {author} {\bibfnamefont {J.~A.}\ \bibnamefont
  {Sherman}}, \bibinfo {author} {\bibfnamefont {N.~B.}\ \bibnamefont
  {Phillips}}, \bibinfo {author} {\bibfnamefont {M.}~\bibnamefont {Schioppo}},
  \bibinfo {author} {\bibfnamefont {N.~D.}\ \bibnamefont {Lemke}}, \bibinfo
  {author} {\bibfnamefont {K.}~\bibnamefont {Beloy}}, \bibinfo {author}
  {\bibfnamefont {M.}~\bibnamefont {Pizzocaro}}, \bibinfo {author}
  {\bibfnamefont {C.~W.}\ \bibnamefont {Oates}}, \ and\ \bibinfo {author}
  {\bibfnamefont {A.~D.}\ \bibnamefont {Ludlow}},\ }\href {\doibase
  10.1126/science.1240420} {\bibfield  {journal} {\bibinfo  {journal}
  {Science}\ }\textbf {\bibinfo {volume} {341}},\ \bibinfo {pages} {1215}
  (\bibinfo {year} {2013})}\BibitemShut {NoStop}%
\bibitem [{\citenamefont {Chen}\ \emph {et~al.}(2011)\citenamefont {Chen},
  \citenamefont {Bohnet}, \citenamefont {Sankar}, \citenamefont {Dai},\ and\
  \citenamefont {Thompson}}]{CBS11}%
  \BibitemOpen
  \bibfield  {author} {\bibinfo {author} {\bibfnamefont {Z.}~\bibnamefont
  {Chen}}, \bibinfo {author} {\bibfnamefont {J.~G.}\ \bibnamefont {Bohnet}},
  \bibinfo {author} {\bibfnamefont {S.~R.}\ \bibnamefont {Sankar}}, \bibinfo
  {author} {\bibfnamefont {J.}~\bibnamefont {Dai}}, \ and\ \bibinfo {author}
  {\bibfnamefont {J.~K.}\ \bibnamefont {Thompson}},\ }\href {\doibase
  10.1103/PhysRevLett.106.133601} {\bibfield  {journal} {\bibinfo  {journal}
  {Phys. Rev. Lett.}\ }\textbf {\bibinfo {volume} {106}},\ \bibinfo {pages}
  {133601} (\bibinfo {year} {2011})}\BibitemShut {NoStop}%
\bibitem [{\citenamefont {Tycko}\ \emph {et~al.}(1985)\citenamefont {Tycko},
  \citenamefont {Cho}, \citenamefont {Schneider},\ and\ \citenamefont
  {Pines}}]{Tycho_composite_1985}%
  \BibitemOpen
  \bibfield  {author} {\bibinfo {author} {\bibfnamefont {R.}~\bibnamefont
  {Tycko}}, \bibinfo {author} {\bibfnamefont {H.~M.}\ \bibnamefont {Cho}},
  \bibinfo {author} {\bibfnamefont {E.}~\bibnamefont {Schneider}}, \ and\
  \bibinfo {author} {\bibfnamefont {A.}~\bibnamefont {Pines}},\ }\href
  {\doibase 10.1016/0022-2364(85)90270-7} {\bibfield  {journal} {\bibinfo
  {journal} {J. Magn. Reson., Ser A}\ }\textbf {\bibinfo {volume} {61}},\
  \bibinfo {pages} {90 } (\bibinfo {year} {1985})}\BibitemShut {NoStop}%
\bibitem [{\citenamefont {Wimperis}(1994)}]{Wimperis_composite_1994}%
  \BibitemOpen
  \bibfield  {author} {\bibinfo {author} {\bibfnamefont {S.}~\bibnamefont
  {Wimperis}},\ }\href {\doibase 10.1006/jmra.1994.1159} {\bibfield  {journal}
  {\bibinfo  {journal} {J. Magn. Reson., Ser A}\ }\textbf {\bibinfo {volume}
  {109}},\ \bibinfo {pages} {221 } (\bibinfo {year} {1994})}\BibitemShut
  {NoStop}%
\bibitem [{\citenamefont {Khaneja}\ \emph {et~al.}(2005)\citenamefont
  {Khaneja}, \citenamefont {Reiss}, \citenamefont {Kehlet}, \citenamefont
  {Schulte-Herbr{\"{u}}ggen},\ and\ \citenamefont
  {Glaser}}]{Khaneja_Composite_2005}%
  \BibitemOpen
  \bibfield  {author} {\bibinfo {author} {\bibfnamefont {N.}~\bibnamefont
  {Khaneja}}, \bibinfo {author} {\bibfnamefont {T.}~\bibnamefont {Reiss}},
  \bibinfo {author} {\bibfnamefont {C.}~\bibnamefont {Kehlet}}, \bibinfo
  {author} {\bibfnamefont {T.}~\bibnamefont {Schulte-Herbr{\"{u}}ggen}}, \ and\
  \bibinfo {author} {\bibfnamefont {S.~J.}\ \bibnamefont {Glaser}},\ }\href
  {\doibase 10.1016/j.jmr.2004.11.004} {\bibfield  {journal} {\bibinfo
  {journal} {J. Magn. Reson., Ser A}\ }\textbf {\bibinfo {volume} {172}},\
  \bibinfo {pages} {296 } (\bibinfo {year} {2005})}\BibitemShut {NoStop}%
\bibitem [{\citenamefont {Li}\ and\ \citenamefont
  {Khaneja}(2006)}]{Li_Composite_2006}%
  \BibitemOpen
  \bibfield  {author} {\bibinfo {author} {\bibfnamefont {J.-S.}\ \bibnamefont
  {Li}}\ and\ \bibinfo {author} {\bibfnamefont {N.}~\bibnamefont {Khaneja}},\
  }\href {\doibase 10.1103/PhysRevA.73.030302} {\bibfield  {journal} {\bibinfo
  {journal} {Phys. Rev. A}\ }\textbf {\bibinfo {volume} {73}},\ \bibinfo
  {pages} {030302} (\bibinfo {year} {2006})}\BibitemShut {NoStop}%
\bibitem [{\citenamefont {Uhrig}(2007)}]{PulseSequence_Uhrig_2007}%
  \BibitemOpen
  \bibfield  {author} {\bibinfo {author} {\bibfnamefont {G.~S.}\ \bibnamefont
  {Uhrig}},\ }\href {\doibase 10.1103/PhysRevLett.98.100504} {\bibfield
  {journal} {\bibinfo  {journal} {Phys. Rev. Lett.}\ }\textbf {\bibinfo
  {volume} {98}},\ \bibinfo {pages} {100504} (\bibinfo {year}
  {2007})}\BibitemShut {NoStop}%
\bibitem [{\citenamefont {Steffen}\ and\ \citenamefont
  {Koch}(2007)}]{Steffen_Composite_2007}%
  \BibitemOpen
  \bibfield  {author} {\bibinfo {author} {\bibfnamefont {M.}~\bibnamefont
  {Steffen}}\ and\ \bibinfo {author} {\bibfnamefont {R.~H.}\ \bibnamefont
  {Koch}},\ }\href {\doibase 10.1103/PhysRevA.75.062326} {\bibfield  {journal}
  {\bibinfo  {journal} {Phys. Rev. A}\ }\textbf {\bibinfo {volume} {75}},\
  \bibinfo {pages} {062326} (\bibinfo {year} {2007})}\BibitemShut {NoStop}%
\bibitem [{\citenamefont {Rakreungdet}\ \emph {et~al.}(2009)\citenamefont
  {Rakreungdet}, \citenamefont {Lee}, \citenamefont {Lee}, \citenamefont
  {Mischuck}, \citenamefont {Montano},\ and\ \citenamefont
  {Jessen}}]{Rakreungdet_Composite_2009}%
  \BibitemOpen
  \bibfield  {author} {\bibinfo {author} {\bibfnamefont {W.}~\bibnamefont
  {Rakreungdet}}, \bibinfo {author} {\bibfnamefont {J.~H.}\ \bibnamefont
  {Lee}}, \bibinfo {author} {\bibfnamefont {K.~F.}\ \bibnamefont {Lee}},
  \bibinfo {author} {\bibfnamefont {B.~E.}\ \bibnamefont {Mischuck}}, \bibinfo
  {author} {\bibfnamefont {E.}~\bibnamefont {Montano}}, \ and\ \bibinfo
  {author} {\bibfnamefont {P.~S.}\ \bibnamefont {Jessen}},\ }\href {\doibase
  10.1103/PhysRevA.79.022316} {\bibfield  {journal} {\bibinfo  {journal} {Phys.
  Rev. A}\ }\textbf {\bibinfo {volume} {79}},\ \bibinfo {pages} {022316}
  (\bibinfo {year} {2009})}\BibitemShut {NoStop}%
\bibitem [{\citenamefont {Torosov}\ and\ \citenamefont
  {Vitanov}(2011)}]{Torosov_Composite_2011}%
  \BibitemOpen
  \bibfield  {author} {\bibinfo {author} {\bibfnamefont {B.~T.}\ \bibnamefont
  {Torosov}}\ and\ \bibinfo {author} {\bibfnamefont {N.~V.}\ \bibnamefont
  {Vitanov}},\ }\href {\doibase 10.1103/PhysRevA.83.053420} {\bibfield
  {journal} {\bibinfo  {journal} {Phys. Rev. A}\ }\textbf {\bibinfo {volume}
  {83}},\ \bibinfo {pages} {053420} (\bibinfo {year} {2011})}\BibitemShut
  {NoStop}%
\bibitem [{\citenamefont {Bohnet}\ \emph {et~al.}(2014)\citenamefont {Bohnet},
  \citenamefont {Cox}, \citenamefont {Norcia}, \citenamefont {Weiner},
  \citenamefont {Chen},\ and\ \citenamefont
  {Thompson}}]{Squeezing_Bohnet_2013}%
  \BibitemOpen
  \bibfield  {author} {\bibinfo {author} {\bibfnamefont {J.~G.}\ \bibnamefont
  {Bohnet}}, \bibinfo {author} {\bibfnamefont {K.~C.}\ \bibnamefont {Cox}},
  \bibinfo {author} {\bibfnamefont {M.~A.}\ \bibnamefont {Norcia}}, \bibinfo
  {author} {\bibfnamefont {J.~M.}\ \bibnamefont {Weiner}}, \bibinfo {author}
  {\bibfnamefont {Z.}~\bibnamefont {Chen}}, \ and\ \bibinfo {author}
  {\bibfnamefont {J.~K.}\ \bibnamefont {Thompson}},\ }\href
  {http://dx.doi.org/10.1038/nphoton.2014.151} {\bibfield  {journal} {\bibinfo
  {journal} {Nat. Photon.}\ }\textbf {\bibinfo {volume} {advance online
  publication}} (\bibinfo {year} {2014})}\BibitemShut {NoStop}%
\bibitem [{\citenamefont {L\"ucke}\ \emph {et~al.}(2014)\citenamefont
  {L\"ucke}, \citenamefont {Peise}, \citenamefont {Vitagliano}, \citenamefont
  {Arlt}, \citenamefont {Santos}, \citenamefont {T\'oth},\ and\ \citenamefont
  {Klempt}}]{PhysRevLett.112.155304}%
  \BibitemOpen
  \bibfield  {author} {\bibinfo {author} {\bibfnamefont {B.}~\bibnamefont
  {L\"ucke}}, \bibinfo {author} {\bibfnamefont {J.}~\bibnamefont {Peise}},
  \bibinfo {author} {\bibfnamefont {G.}~\bibnamefont {Vitagliano}}, \bibinfo
  {author} {\bibfnamefont {J.}~\bibnamefont {Arlt}}, \bibinfo {author}
  {\bibfnamefont {L.}~\bibnamefont {Santos}}, \bibinfo {author} {\bibfnamefont
  {G.}~\bibnamefont {T\'oth}}, \ and\ \bibinfo {author} {\bibfnamefont
  {C.}~\bibnamefont {Klempt}},\ }\href {\doibase
  10.1103/PhysRevLett.112.155304} {\bibfield  {journal} {\bibinfo  {journal}
  {Phys. Rev. Lett.}\ }\textbf {\bibinfo {volume} {112}},\ \bibinfo {pages}
  {155304} (\bibinfo {year} {2014})}\BibitemShut {NoStop}%
\bibitem [{\citenamefont {Hamley}\ \emph {et~al.}(2012)\citenamefont {Hamley},
  \citenamefont {Gerving}, \citenamefont {Hoang}, \citenamefont {Bookjans},\
  and\ \citenamefont {Chapman}}]{ChapmanSqueeze}%
  \BibitemOpen
  \bibfield  {author} {\bibinfo {author} {\bibfnamefont {C.~D.}\ \bibnamefont
  {Hamley}}, \bibinfo {author} {\bibfnamefont {C.~S.}\ \bibnamefont {Gerving}},
  \bibinfo {author} {\bibfnamefont {T.~M.}\ \bibnamefont {Hoang}}, \bibinfo
  {author} {\bibfnamefont {E.~M.}\ \bibnamefont {Bookjans}}, \ and\ \bibinfo
  {author} {\bibfnamefont {M.~S.}\ \bibnamefont {Chapman}},\ }\href
  {http://dx.doi.org/10.1038/nphys2245} {\bibfield  {journal} {\bibinfo
  {journal} {Nat. Phys.}\ }\textbf {\bibinfo {volume} {8}},\ \bibinfo {pages}
  {305} (\bibinfo {year} {2012})}\BibitemShut {NoStop}%
\bibitem [{\citenamefont {Monz}\ \emph {et~al.}(2011)\citenamefont {Monz},
  \citenamefont {Schindler}, \citenamefont {Barreiro}, \citenamefont {Chwalla},
  \citenamefont {Nigg}, \citenamefont {Coish}, \citenamefont {Harlander},
  \citenamefont {H\"ansel}, \citenamefont {Hennrich},\ and\ \citenamefont
  {Blatt}}]{BlattSqueeze}%
  \BibitemOpen
  \bibfield  {author} {\bibinfo {author} {\bibfnamefont {T.}~\bibnamefont
  {Monz}}, \bibinfo {author} {\bibfnamefont {P.}~\bibnamefont {Schindler}},
  \bibinfo {author} {\bibfnamefont {J.~T.}\ \bibnamefont {Barreiro}}, \bibinfo
  {author} {\bibfnamefont {M.}~\bibnamefont {Chwalla}}, \bibinfo {author}
  {\bibfnamefont {D.}~\bibnamefont {Nigg}}, \bibinfo {author} {\bibfnamefont
  {W.~A.}\ \bibnamefont {Coish}}, \bibinfo {author} {\bibfnamefont
  {M.}~\bibnamefont {Harlander}}, \bibinfo {author} {\bibfnamefont
  {W.}~\bibnamefont {H\"ansel}}, \bibinfo {author} {\bibfnamefont
  {M.}~\bibnamefont {Hennrich}}, \ and\ \bibinfo {author} {\bibfnamefont
  {R.}~\bibnamefont {Blatt}},\ }\href {\doibase 10.1103/PhysRevLett.106.130506}
  {\bibfield  {journal} {\bibinfo  {journal} {Phys. Rev. Lett.}\ }\textbf
  {\bibinfo {volume} {106}},\ \bibinfo {pages} {130506} (\bibinfo {year}
  {2011})}\BibitemShut {NoStop}%
\bibitem [{\citenamefont {Leroux}\ \emph {et~al.}(2010)\citenamefont {Leroux},
  \citenamefont {Schleier-Smith},\ and\ \citenamefont {Vuleti\'{c}}}]{LSM10}%
  \BibitemOpen
  \bibfield  {author} {\bibinfo {author} {\bibfnamefont {I.~D.}\ \bibnamefont
  {Leroux}}, \bibinfo {author} {\bibfnamefont {M.~H.}\ \bibnamefont
  {Schleier-Smith}}, \ and\ \bibinfo {author} {\bibfnamefont {V.}~\bibnamefont
  {Vuleti\'{c}}},\ }\href {\doibase 10.1103/PhysRevLett.104.073602} {\bibfield
  {journal} {\bibinfo  {journal} {Phys. Rev. Lett.}\ }\textbf {\bibinfo
  {volume} {104}},\ \bibinfo {pages} {073602} (\bibinfo {year}
  {2010})}\BibitemShut {NoStop}%
\bibitem [{Note1()}]{Note1}%
  \BibitemOpen
  \bibinfo {note} {A different measurement sequence (Ref. \protect \citenum
  {Squeezing_Bohnet_2013}) that avoids any rotations achieved a comparable
  amount of directly observed spin squeezing in the same system without relying
  on dephased rotations, but only by sacrificing a factor of two in fundamental
  measurement resolution. This additional factor of two was not realized in
  this work due to the probe laser's frequency noise coupling more strongly
  into the measurment sequence of Fig.~\ref {fig:fig3}}\BibitemShut {NoStop}%
\bibitem [{\citenamefont {Chen}\ \emph {et~al.}(2012)\citenamefont {Chen},
  \citenamefont {Bohnet}, \citenamefont {Weiner},\ and\ \citenamefont
  {Thompson}}]{PhaseNoise_Chen_2012}%
  \BibitemOpen
  \bibfield  {author} {\bibinfo {author} {\bibfnamefont {Z.}~\bibnamefont
  {Chen}}, \bibinfo {author} {\bibfnamefont {J.~G.}\ \bibnamefont {Bohnet}},
  \bibinfo {author} {\bibfnamefont {J.~M.}\ \bibnamefont {Weiner}}, \ and\
  \bibinfo {author} {\bibfnamefont {J.~K.}\ \bibnamefont {Thompson}},\ }\href
  {\doibase 10.1103/PhysRevA.86.032313} {\bibfield  {journal} {\bibinfo
  {journal} {Phys. Rev. A}\ }\textbf {\bibinfo {volume} {86}},\ \bibinfo
  {pages} {032313} (\bibinfo {year} {2012})}\BibitemShut {NoStop}%
\bibitem [{\citenamefont {Chen}\ \emph {et~al.}(2014)\citenamefont {Chen},
  \citenamefont {Bohnet}, \citenamefont {Weiner}, \citenamefont {Cox},\ and\
  \citenamefont {Thompson}}]{PhysRevA.89.043837}%
  \BibitemOpen
  \bibfield  {author} {\bibinfo {author} {\bibfnamefont {Z.}~\bibnamefont
  {Chen}}, \bibinfo {author} {\bibfnamefont {J.~G.}\ \bibnamefont {Bohnet}},
  \bibinfo {author} {\bibfnamefont {J.~M.}\ \bibnamefont {Weiner}}, \bibinfo
  {author} {\bibfnamefont {K.~C.}\ \bibnamefont {Cox}}, \ and\ \bibinfo
  {author} {\bibfnamefont {J.~K.}\ \bibnamefont {Thompson}},\ }\href {\doibase
  10.1103/PhysRevA.89.043837} {\bibfield  {journal} {\bibinfo  {journal} {Phys.
  Rev. A}\ }\textbf {\bibinfo {volume} {89}},\ \bibinfo {pages} {043837}
  (\bibinfo {year} {2014})}\BibitemShut {NoStop}%
\bibitem [{\citenamefont {Kitagawa}\ and\ \citenamefont
  {Ueda}(1993)}]{kitagawa_1993}%
  \BibitemOpen
  \bibfield  {author} {\bibinfo {author} {\bibfnamefont {M.}~\bibnamefont
  {Kitagawa}}\ and\ \bibinfo {author} {\bibfnamefont {M.}~\bibnamefont
  {Ueda}},\ }\href {\doibase 10.1103/PhysRevA.47.5138} {\bibfield  {journal}
  {\bibinfo  {journal} {Phys. Rev. A}\ }\textbf {\bibinfo {volume} {47}},\
  \bibinfo {pages} {5138} (\bibinfo {year} {1993})}\BibitemShut {NoStop}%
\end{thebibliography}%

\end{document}